\documentclass[prl,twocolumn,amsmath]{revtex4-1}

\usepackage{amsmath}
\usepackage{amssymb}
\usepackage{color}
\usepackage[normalem]{ulem}
\usepackage{mathtools}
\usepackage{color}

\begin{document}
\title{From Quenched Disorder to Continuous Time Random Walk }
\author{Stanislav Burov}
\email{stasbur@gmail.com}
\affiliation{Physics Department, Bar-Ilan University, Ramat Gan 5290002,
Israel}

\pacs{PACS}

\begin{abstract}
 This work focuses on quantitative representation of transport in systems with quenched disorder.
 Explicit mapping of the quenched trap model to  continuous time random walk is presented. Linear temporal transformation: $t\to t/\Lambda^{1/\alpha}$ for transient process 
 in the sub-diffusive regime, is sufficient for asymptotic mapping. 
 Exact form of the constant $\Lambda^{1/\alpha}$ is established.
 Disorder averaged position probability density function for quenched trap model is obtained and analytic expressions for the diffusion coefficient  and drift are provided.
\end{abstract}

\maketitle

Properties of transport in disordered environment are objects of intensive research~\cite{Alexander,Bouchaud,Klafter}. 
While regular diffusion is vastly observed in many systems, anomalously slow diffusion (i.e. $\langle x^2(t) \rangle \sim t^ \alpha$ where $0<\alpha<1$) effectively describes motion in complex disordered systems such as living cells~\cite{LiveCell,Tabei}, blinking quantum dots~\cite{QuantumD}, molecular-motor transport on filament network~\cite{Cycling} and photo-currents in amorphous materials~\cite{ScherMontroll}. 
Several theoretical approaches give rise to anomalous diffusion of a particle in disordered media. The Fractional Brownian Motion~\cite{Mandelbrot} effectively models the disorder as long-ranged temporal correlations. Another approach attributes the slow-down to presence of obstacles, such as traps and barriers, in the media. For example, random walks (RW) obstructed by traps~\cite{LiveCell,Cycling} and barriers~\cite{Bariers} were used to model properties of intracellular transport. When the expected local dwell times diverge, the diffusion becomes anomalous~\cite{Bouchaud, Klafter}.

Transport mediated by traps and barriers attracted tremendous attention in Physics and Mathematics. The usual theoretical description consists of a RW on a lattice, where the disorder enters via transition probabilities (and rates) to different lattice sites. 
Two general disorder types prevail: annealed disorder and quenched disorder. The annealed disorder describes the situation when the disorder is uncorrelated. For each visit to a lattice site  new disorder is generated. On the contrary, quenched disorder suggests that the disorder per site stays exactly the same for all visits of the RW. 
This imposes strong correlations and makes theoretical description highly non-trivial. When using traps as  disorder, the dwell time at  specific lattice site can be constant (quenched) or generated from a random distribution for each arrival (annealed). The later model is known as continuous time random walk (CTRW)~\cite{ScherMontroll} and its behavior is well known~\cite{Bouchaud,Weiss}. Once the dwell times are quenched, a case known as the quenched trap model (QTM), the renewal property is lost. 
Scaling arguments and renormalization group approach~\cite{Bouchaud, Machta, Monthus} suggest that for dimension $d>2$ QTM behaves qualitatively as CTRW in the sub-diffusive phase. Similar result was suggested by using  rigorous mathematical description of QTM on a regular lattice~\cite{BenArous1,BenArous2}. Simple hand-waving argument behind this convergence is based on the fact that for $d>2$ the probability of RW to return to a specific site is $<1$. One can then assert that the correlations imposed by quenched dwell times can be effectively renormalized into uncorrelated times, i.e. CTRW description. 
Similar argument should also hold for the case of a biased transport, i.e. RW with directional preference. For example the case of directional RW (transitions only in one direction) for QTM in $d=1$~\cite{Aslangul}  is believed to be asymptotically similar to the general biased case. While for directed RW the particle never returns to the same site, for a general biased case the probability of return is $<1$.

In this manuscript explicit mapping between QTM (quenched disorder) and CTRW (annealed disorder) is provided. It will be shown that for any case when a RW is transient, i.e. the probability of return $<1$,  the probability density function (PDF) in the sub-diffusive regime takes the form of an appropriate CTRW process. The missing {\it{quantitative}} representation of QTM in terms of CTRW will be provided for any QTM that takes place on translationally invariant lattice. Transitions between different lattice sites are not restricted to nearest neighbors.   The presented approach  is based on reformulation of subordination technique for CTRW~\cite{Bouchaud,Fogedby, Barkai} that was introduced in ~\cite{Burov1,Burov2}.

{\it{The Quenched Trap Model}} is defined as a random process on a lattice of dimension $d$. For each site $\bf{x}$ of the lattice a quenched random variable $\tau_{\bf{x}}$ is defined. $\tau_{\bf{x}}$ describes the time that the particle spends at site $\bf{x}$ before moving to some random site ${\bf{x}}' $. 
The process starts at time $t=0$ when the particle is situated at ${\bf{x}}=0$. Probability of transition from ${\bf{x}}$ to ${\bf{x}}'$ is provided by $p({\bf{x}}';{\bf{x}})$. Due to translational invariance of the lattice $p({\bf{x}}';{\bf{x}})$ is a function of ${\bf{x}}'-{\bf{x}}$, i.e $p({\bf{x}}'-{\bf{x}})$. The quenched variables $\{ \tau_{\bf{x}}\}$ are positive, independent and identically distributed random variables with common PDF 
$\psi(\tau_{\bf{x}})\sim \tau_{\bf{x}}^{-(1+\alpha)} A/|\Gamma(-\alpha)|$ for $\tau_{\bf{x}}\to\infty$ ($A>0$ and $\Gamma(\dots)$ is the Gamma function). The values of $\alpha$ will be restricted to $0<\alpha<1$ in order to describe the subdiffusve regime of QTM~\cite{Bouchaud}. Local dwell times $\tau_{\bf{x}}$ describe for how long the particle is "trapped" on site ${\bf{x}}$. The physical picture is usually attributed to thermally activated jumps upon random energy potential. Each lattice site is associated with energetic trap with energy depth $E>0$ that is exponentially distributed, i.e. $f(E)=\exp(-E/T_g)/T_g$.

One thing to notice about QTM is that if the process is observed as a function of number of performed steps, it behaves like a  RW with transition probabilities defined by $p({\bf{x}}'-{\bf{x}})$. 
Similar statement is true for CTRW. The ``solution" of QTM is then a proper  transformation from the number of steps to ordinary time. Time is a function of all possible traps that the particle encountered on its path. In QTM time is provided by $t=\sum_{\bf{x}}n_{\bf{x}}\tau_{\bf{x}}$, where $n_{\bf{x}}$ is the number of visits to site ${\bf{x}}$. The sum follows all  different sites on the lattice.
Similarly to~\cite{Burov1,Burov2} a random variable $S_\alpha$ is defined
\begin{equation}
S_\alpha=\sum_{\bf{x}} (n_{\bf{x}})^\alpha,
\label{salphadef}
\end{equation}
 and the sum is again over all lattice sites. $S_\alpha$ is a spatial variable which depends solely on various positions of the particle and not the time spent at those sites. For $\alpha=1$ $S_\alpha$ is the total number of steps performed. 
 In~\cite{Burov1} it was shown that the random variable $\eta=t/\left( S_{\bf{x}} \right)^{1/\alpha}$ is distributed according to one-sided L\'{e}vy PDF $l_{\alpha,A,1}(\eta)$~\cite{Barkai}. The argument is as follows: while averaging the quantity $\exp(-\eta u)$  ($u>0$) over disorder, it occurs that
 \begin{equation}
{\Big{\langle}} e^{-\eta u}{\Big{\rangle}}={\Big{\langle}} \exp\left( -\sum_{\bf{x}}\frac{n_{\bf{x}}\tau_{\bf{x}}}{S_\alpha^{1/\alpha}} \right){\Big{\rangle}}\to e^{-Au^{\alpha}}
 \label{limitLap}
 \end{equation}
and $e^{-Au^{\alpha}}$ is the Laplace pair of $l_{\alpha,A,1}(\eta)$. When constraining $t$ to a fixed value, the PDF of $S_\alpha$ is easily obtained from the definition of $\eta$
\begin{equation}
{\cal{N}} _t (S_\alpha)=\frac{t}{\alpha}\left( S_\alpha \right)^{-\frac{1}{\alpha}-1} l_{\alpha,A,1} 
\left(\frac{t}{S_\alpha ^{1/\alpha}} \right).
\label{distributionS}
\end{equation}
Equation~(\ref{distributionS}) defines the distribution of $S_\alpha$ and is a part of transformation from accumulated disorder to real time. The probability of arriving to ${\bf{x}}$ at time $t$ can be separated into probability of arriving to ${\bf{x}}$ at some $S_\alpha$ and probability of observing this specific $S_\alpha$, i.e. ${\cal{N}}_t(S_\alpha)$. $S_\alpha$ is  operational time of the process and Eq.~(\ref{distributionS}) is the transformation from operational time to real time $t$. 
For specific $S_\alpha$ the probability to observe the particle at ${\bf{x}}$ for specific $S_\alpha$  is written as $P_{S_\alpha}({\bf{x}})$. 
Disorder averaged PDF of position ${\bf{x}}$ at time $t$  is provided by 
$\langle P({\bf{x}},t) \rangle=\sum_{S_\alpha} P_{S_\alpha} ({\bf{x}}) {\cal{N}} _t (S_\alpha)$, where the sum is over all possible $S_\alpha$s. Notice that $ P_{S_\alpha}$ is independent of disorder. $S_\alpha$ is positively defined and $\langle P({\bf{x}},t) \rangle$ is written as
\begin{equation}
\langle P({\bf{x}},t) \rangle\sim\int_0^\infty P_{S_\alpha} ({\bf{x}}) {\cal{N}} _t (S_\alpha)\,dS_\alpha.
\label{transformG}
\end{equation}
Since $ {\cal{N}} _t (S_\alpha)$ is given by Eq.~(\ref{distributionS}) the problem of determining $\langle P({\bf{x}},t) \rangle$ for QTM boils down to determining $P_{S_\alpha} ({\bf{x}})$, which is a property of RW on a lattice. 
  
\begin{figure}[h!]
\begin{center}

                \includegraphics[width=0.49\textwidth]{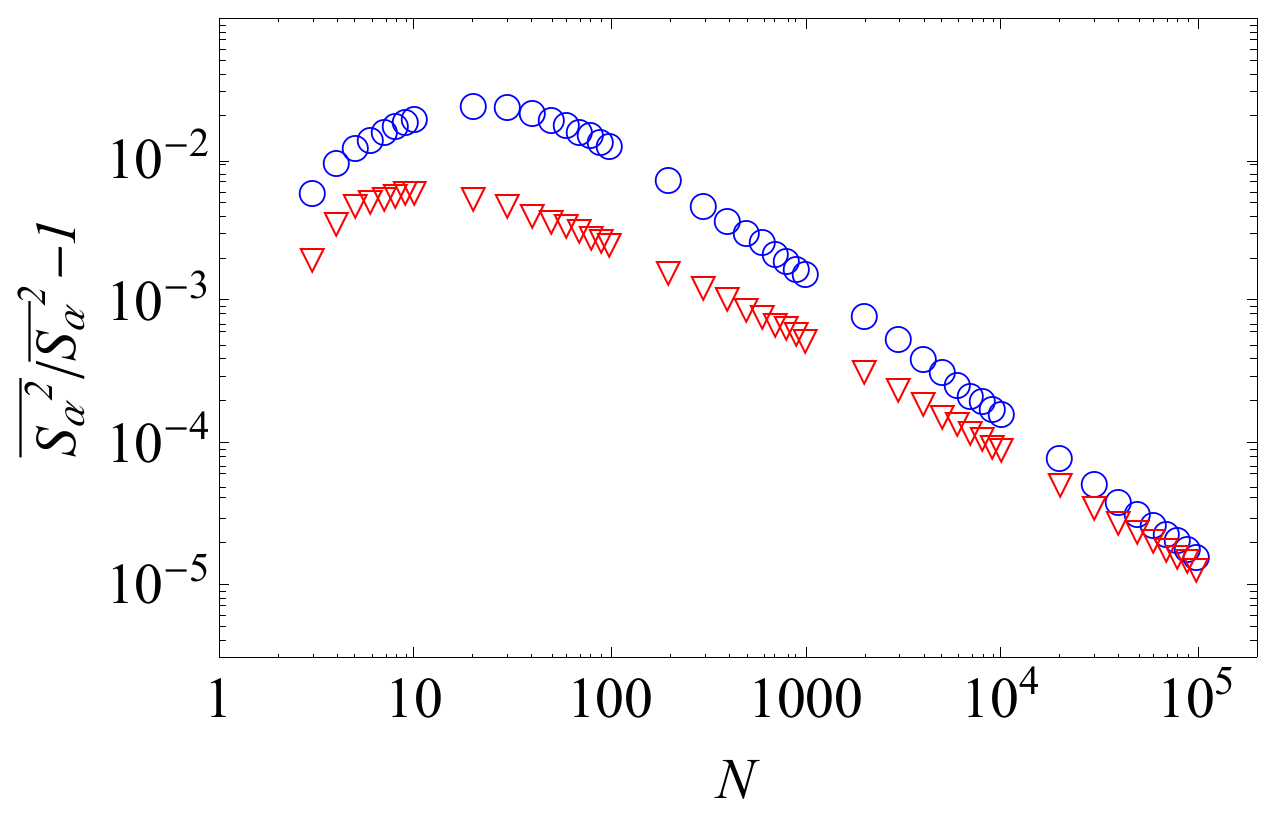}

\end{center}
\caption{Simulated behavior of fraction of moments of $S_\alpha$, i.e. ${\overline{S_\alpha^2}}\Big/{\overline{S_\alpha^2}}^2-1$, as function of the number of jumps  ($N$) of a random walk. {\color{blue}${\bigcirc}$} are the results for a biased one-dimensional RW on a lattice, the transitions are allowed only to nearest neighbors with probability $q=0.7$ to the right and $1-q$ to the left.  {\color{red}$\bigtriangledown$} presents the results for $3$-dimensional unbiased RW on a cubic lattice where the transitions are allowed only to nearest neighbors.
}
\label{figsalpha}
\end{figure}

Although operational time $S_\alpha$ is defined for a RW without disorder its behavior is quite non-trivial since it is defined by the whole history of a random trajectory. 
$P_{S_\alpha} ({\bf{x}})$ describes a random walk that was stopped at specific $S_\alpha$ while the number of performed steps is arbitrary. In~\cite{Burov1} it was shown that for $d=1$ and nearest-neighbor jumps of the RW, $P_{S_\alpha} ({\bf{x}})$ attains transition from a Gaussian shape ($\alpha\to1$) to a $V$ shape ($\alpha\to0$). 
It is the purpose of this manuscript to show that for any transient RW, $P_{S_\alpha} ({\bf{x}})$ is easily obtained from 
${\cal{P}}_N({\bf{x}})$, i.e. the probability to find the particle at position ${\bf{x}}$ after $N$ steps.
\cite{BurovF} will provide a mathematical proof that for transient RW (on translationally invariant lattice) the fraction of the moments of $S_\alpha$, i.e. $\overline {S_\alpha^2}/{\overline{S_\alpha}}^2$, converges to $1$ as $N\to\infty$.  The average ($\overline{\quad}$) is taken with respect to all possible RW that start at the origin and perform $N$ steps. 
In Fig.~\ref{figsalpha} the convergence of fraction of moments is presented for two different cases of transient RWs.
It is shown below that in the limit of large $N$, ${\overline{S_\alpha}}/N$ converges to a non-zero constant. Since ${\overline {S_\alpha^2}}/{\overline{S_\alpha}}^2\to1$, it means that $S_\alpha/N$ converges to a $\delta$-function.
 By calculation of ${\overline{S_\alpha}}$ the {\bf{deterministic mapping}} between $S_\alpha$ and $N$ is found. 
This mapping determines $N$ as a function of $S_\alpha$, i.e. $N(S_\alpha)$, and consequently $P_{S_\alpha}({\bf{x}})\sim{\cal P}_{N(S_\alpha)} ({\bf{x}})$.  Since ${\cal P}_N({\bf{x}})$ describes RW on a spatially invariant lattice, its properties are well documented~\cite{Weiss}.

{\it Calculation of ${\overline{S_\alpha}}$}. Let $\beta_N({\bf{x}};k)$ be a probability that a RW visited site ${\bf{x}}$ exactly $k$ times after $N$ steps. ${\overline{S_\alpha}}$ is expressed in terms of $\beta_N({\bf{x}};k)$ as ${\overline{S_\alpha}}=\sum_{\bf{x}}\sum_{k=0}^{k=\infty} k^\alpha\beta_N({\bf{x}};k)$. A closely related quantity is $V_N(k)$, the average number of lattice sites visited exactly $k$ times after $N$ steps. $V_N(1)$ was first derived in~\cite{Erdos}  and for general $k$  using the generating function approach~\cite{Weiss}. The derivation below follows~\cite{Weiss}. By virtue of $f_N({\bf{0}})$, the probability of first return to ${\bf{x}}={\bf{0}}$ after $N$ steps,  we write $f_N({\bf{x}};k)$, the probability to reach site ${\bf{x}}$ for $k$'th time after $N$ steps,  as:  $f_N({\bf{x}};k+1)=\sum_{m=0}^N f_m({\bf{x}};k) f_{N-m}({\bf{0}})$. This relation holds for any translationally invariant lattice.
The generating function of $f_N({\bf{x}};k)$, ${\hat f}_z({\bf{x}};k):=\sum_{N=0}^{\infty} z^N f_N({\bf{x}};k)$, is 
\begin{equation}
{\hat f}_z({\bf{x}};k)=\left[{\hat f}_z({\bf{0}}) \right]^{k-1} {\hat{f}}_z({\bf{x}}),
\label{genFP}
\end{equation}
where ${\hat f}_z({\bf{0}})$ is the generating function of $f_N({\bf{0}})$ and ${\hat f}_z({\bf{x}})$ is the generating function of $f_N({\bf{x}})$ (the probability of first arrival to ${\bf{x}}$).
Since RW must arrive to site ${\bf{x}}$ for $k$th time after $m\leq N$ step (and afterwords can't visit again) $\beta_N({\bf{x}};k)$ takes the form
\begin{equation}
\begin{split}
&\beta_N({\bf{x}};k)=\sum_{m=1}^{N}\left[f_m({\bf{x}};k)-f_m({\bf{x}};k+1) 
\right]\quad\quad{\bf{x}}\neq{\bf{0}}
\\
&\beta_N({\bf{0}};k)=\sum_{m=1}^{N}\left[f_m({\bf{0}};k-1)-f_m({\bf{0}};k) 
\right]
\end{split}
\label{beta1}
\end{equation}
By taking  $z$-transform of both sides in Eq.~(\ref{beta1}) and applying Eq.~(\ref{genFP}),  generating function of $\beta_N({\bf{x}};k)$ is obtained
\begin{equation}
\begin{split}
&{\hat \beta}_z({\bf{x}};k)=\frac{1}{1-z}\left[1-{\hat f}_z({\bf{0}}) \right] \left[ {\hat f}_z({\bf{0}})\right]^{k-1} {\hat f}_z({\bf{x}})
\quad\quad {\bf{x}}\neq{\bf{0}}
\\
&
{\hat \beta}_z({\bf{0}};k)=\frac{1}{1-z}\left[1-{\hat f}_z({\bf{0}}) \right] \left[ {\hat f}_z({\bf{0}})\right]^{k-1}. 
\end{split}
\label{beta2}
\end{equation}
Since ${\cal P}_N({\bf{x}})$ can be written in terms of $f_N({\bf{x}})$, ${\cal P}_N({\bf{x}})=\delta_{N,0}\delta_{{\bf{x}},{\bf{0}}}+\sum_{m=1}^N f_k({\bf{x}}) {\cal P}_{N-m}({\bf{0}})$~\cite{Kronecker}, a known~\cite{Redner} relation holds for generating functions of ${\cal P}_N({\bf{x}})$ and $f_N({\bf{x}})$, i.e., ${\hat f}_z({\bf{x}}\neq{\bf{0}})={\hat{\cal P}}_z({\bf{x}}\neq{\bf{0}})/{\hat{\cal P}}_z({\bf{0}})$; ${\hat f}_z({\bf{0}})=1-1/{\hat{\cal P}}_z({\bf{0}})$. Using these expressions, and the fact that $\sum_{\bf{x}}{\hat{\cal P}}_z({\bf{x}})=1/(1-z)$, we obtain for the generating function of averaged operational time  
\begin{equation}
{\Hat{{\overline{S_\alpha}}}}(z) = \sum_{k=0}^{\infty}k^\alpha\left[\frac{1-{\hat f}_z({\bf{0}})}{1-z}\right]^2 {\hat f}_z({\bf{0}})^{k-1}.
\label{salphaZ}
\end{equation}
${\hat f}_z({\bf{0}})$ is related to $Q_0$, the probability of a RW to return to the origin, since $Q_0=\sum_{N=0}^\infty f_N({\bf{0}})$. By taking the $z\to1$ limit and applying Tauberian theorem~\cite{Feller}, Eq.~(\ref{salphaZ}) is transformed to
\begin{equation}
{\overline{S_\alpha}}\sim \Lambda N 
\quad\quad \left(N\to\infty\right)
\label{avrgsalpha}
\end{equation}
where 
\begin{equation}
\Lambda =  \frac{\left[ 1-Q_0 \right]^2}{Q_0}{\text{Li}}_{-\alpha}\left(Q_0\right)
\label{lambdaDef}
\end{equation}
and ${\text{Li}}_a(b)=\sum_{k=0}^\infty b^k/k^a$ is the Polylogarithm function. Eq.~(\ref{avrgsalpha}) holds in the asymptotic limit of large number of steps and only for $Q_0<1$, i.e. transient RW. 
The linear relation between $N$ and ${\overline{S_\alpha}}$, together with the convergence of $S_\alpha/N$ to a constant value ~\cite{BurovF}, enables us to establish the mapping between QTM and CTRW.

\begin{figure}[h!]
\begin{center}

                \includegraphics[width=0.42\textwidth]{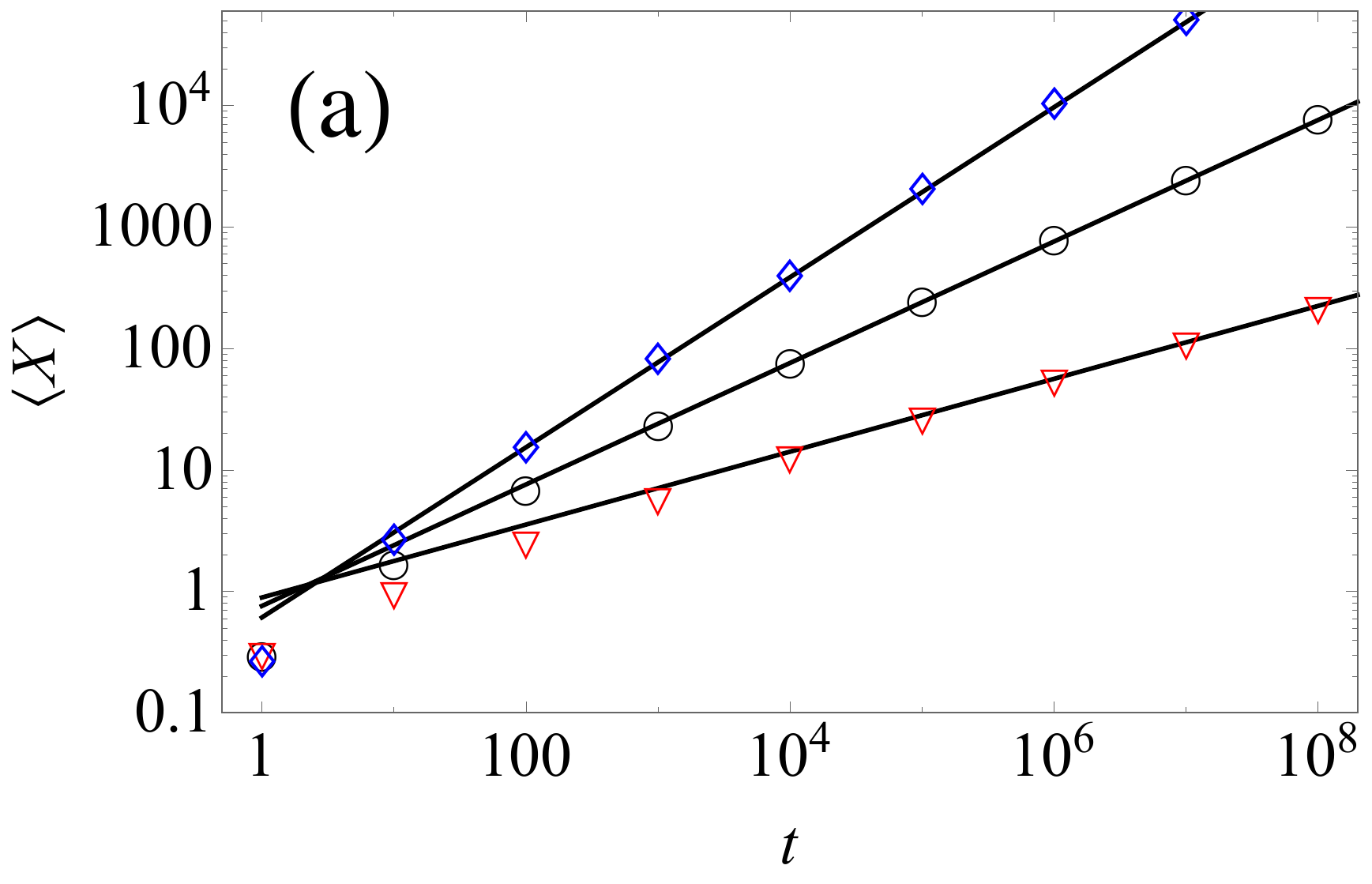}

                \includegraphics[width=0.42\textwidth]{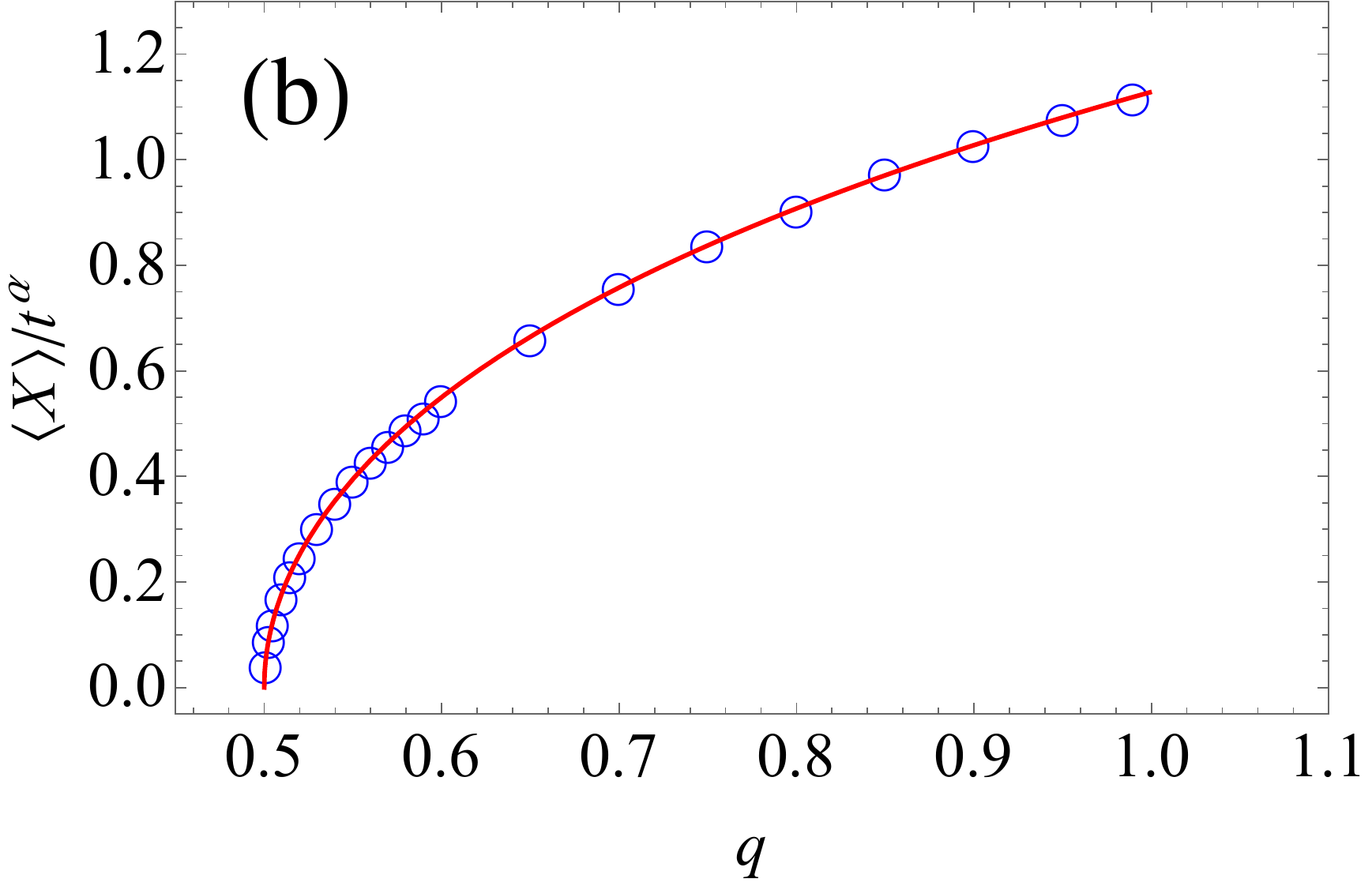}

                \includegraphics[width=0.42\textwidth]{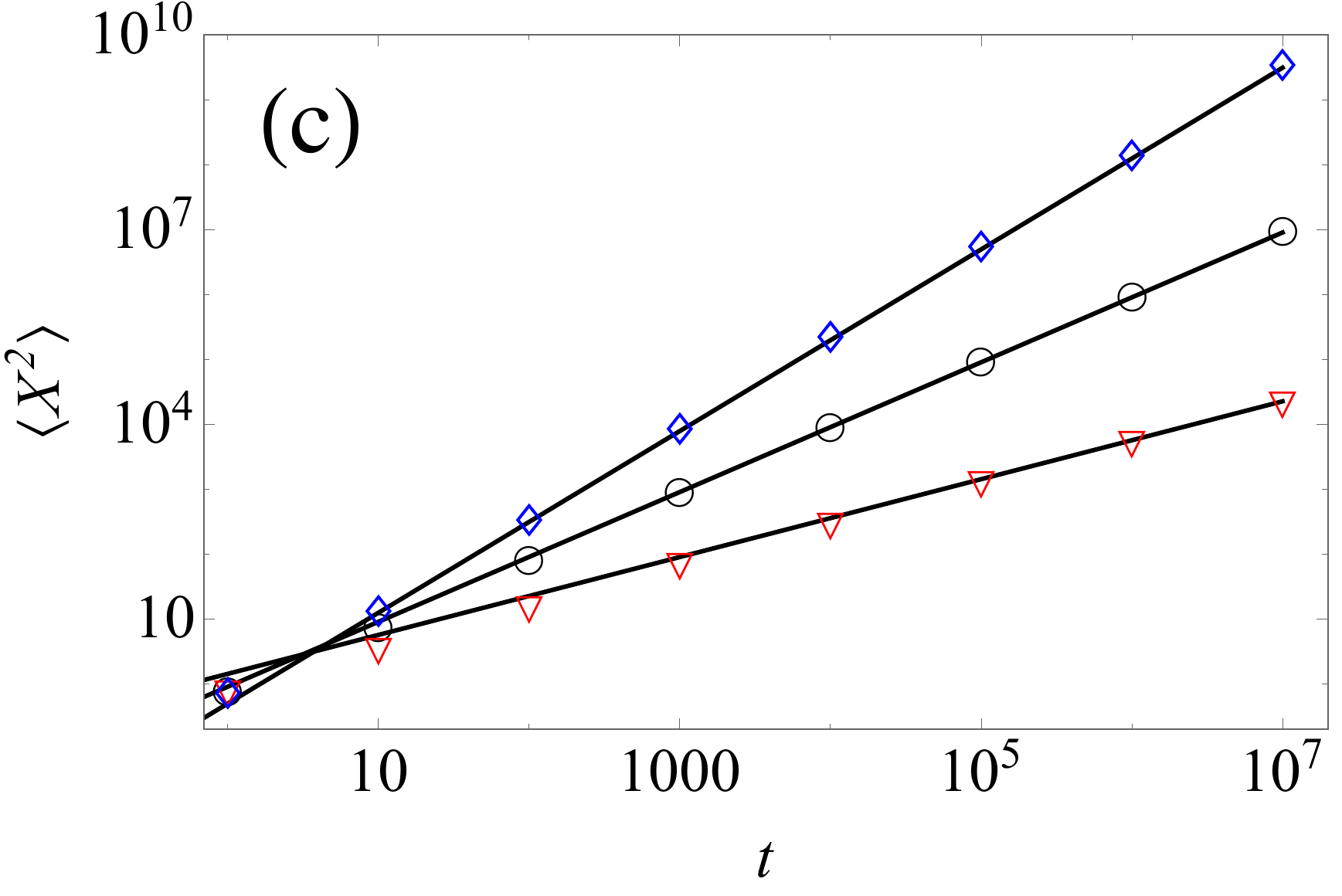}    
\end{center}
\caption{ Moments and pre-factors for $1$-dimensional biased QTM as presented by Eq.~(\ref{biasmoment}) (lines) and numerical simulations (symbols). {\bf{(a)}} presents the first moment behavior as a function of time for three different $\alpha$: {\color{blue}$\diamondsuit$} is $\alpha=0.7$, $\bigcirc$ is $\alpha=0.5$ and {\color{red}$\bigtriangledown$} is $\alpha=0.3$. {\bf{(b)}} presents the behavior of pre-factor $\langle X \rangle/t^\alpha=V/(A\Gamma[1+\alpha])$ for $\alpha=0.5$ and various $q>0.5$. Simulations were performed up to $t=10^7$.  {\bf{(c)}} presents the growth of the second moment with time, the parameters similar to {\bf{(a)}}.
$10^4$ realizations of disorder were used for averaging and $A=1$.
}
\label{figbias}
\end{figure}

{\it{Asymptotic mapping to CTRW}}. $\langle P({\bf{x}};t) \rangle  $ behavior in the asymptotic limit $t\to\infty$ is achieved by substituting ${\cal P}_{N({\overline{S_\alpha}})}({\bf{x}})$ into Eq.~(\ref{transformG}) instead of $P_{{S_\alpha}}({\bf{x}})$.  
The  $t\to\infty$ regime makes sure, by the means of ${\cal N}_t(S_\alpha)$, that sufficient amount of steps has been performed and ${\overline{S_\alpha}}\sim S_\alpha$. Further, a change of variables in Eq.~(\ref{transformG}), 
$S_\alpha\to\Lambda\nu$, leads to 
\begin{equation}
\langle P({\bf{x}};t)\rangle\sim\int_0^\infty {\cal P}_\nu({\bf{x}}) \frac{[t\big/\Lambda^{1/\alpha}]}{\alpha\, \nu ^{-(\frac{1}{\alpha}+1)}} l_{\alpha,A,1}
\left(\frac{[t\big/\Lambda^{1/\alpha}]}{\nu ^{1/\alpha}} \right)d\nu.
\label{pxtfull}
\end{equation}
For CTRW there are no correlations between different waiting times and each site is considered as a new one, from the dwell time perspective. The operational time $S_\alpha$ for CTRW is then simply $N$ and the position PDF is provided by Eq.~(\ref{transformG})~\cite{Bouchaud,Barkai}. From Eq.~(\ref{pxtfull}), and the mentioned representation of CTRW, follows that
\begin{equation}
\Big\langle P({\bf{x}};t) \Big\rangle_{\text{QTM}}\sim
\Big\langle P({\bf{x}};t\big/ \Lambda^{1/\alpha}) \Big\rangle_{\text{CTRW}}
\quad \left(t\to\infty\right),
\label{mapPdf}
\end{equation}
where $\langle\dots\rangle_{\text{QTM}}$ means averaging with respect to quenched disorder of QTM 
and $\langle\dots\rangle_{\text{CTRW}}$ is averaging with respect to annealed disorder of CTRW. 
Eq.~(\ref{mapPdf}) is the main result of this manuscript, simple linear time transformation, $t\to t/\Lambda^{1/\alpha}$, between quenched and annealed disorder. The immediate outcome  is that many known results for CTRW are naturally transformed to quantitative results for QTM. The only limitation of the transformation is the transience of the spatial RW ($Q_{\bf{0}}<1$).

Computation of different positional moments, i.e., $\langle {\bf{x}}^\mu(t) \rangle =\int_{\bf{x}} {\bf{x}}^\mu\Big\langle P({\bf{x}};t) \Big\rangle\,d{\bf{x}}$, becomes quite straightforward in the long time limit. Indeed, by application of Eq.~(\ref{pxtfull}) the spatial integration is preformed only for ${\bf{x}}^\mu{\cal P}_\nu({\bf{x}})$. In the limit of large $\nu$, $\int_{\bf{x}}{\bf{x}}^\mu{\cal P}_\nu({\bf{x}})\,d{\bf{x}}\sim{{\bf{\cal B}}}_\mu \nu^{\gamma_\mu}$. We use $\int_0^\infty y^{q} l_{\alpha,1,1}(y)\,dy=\Gamma(1-q/\alpha)/\Gamma(1-q)$ (for $q/\alpha<1$) and obtain
\begin{equation}
\langle {\bf{x}}^\mu(t) \rangle \sim {{\bf{\cal B}}}_\mu\frac{\Gamma[1+\gamma_\mu]}{\Gamma[1+\alpha\gamma_\mu]}
\left(\frac{Q_0}{A [1-Q_0]^2 {\text{Li}}_{-\alpha}\left(Q_0\right)}\right)^{\gamma_\mu} t^{\alpha\gamma_\mu}.
\label{momentGen}
\end{equation}
Constants ${{\bf{\cal B}}}_\mu$, $\gamma_\mu$ and $Q_0$ depend only on the lattice type and transition probabilities $p({\bf{x}})$. Since the calculation is performed for large times, ${\cal P}_{\nu}({\bf{x}})$ usually converges to Gaussian or L\'{e}vy distribution~\cite{Klafter} where all the moments and pre-factors like ${{\bf{\cal B}}}_\mu$ are known. By the same token, or by simpler scaling arguments, the exponent $\gamma_\mu$ can be obtained. Return probability $Q_0$ has been successfully computed for quite a long time ago~\cite{Watson} for various lattices, in Appendix of \cite{Hughes} (and references therein) appear numerous exact values for $Q_0$.  Two examples of moment behavior are in place (i) biased RW on symmetric lattice in $d=1$ and (ii) non-biased RW on a cubic lattice ($d=3$).

The biased RW in $1$-dimension can perform a unit step to the right with probability $q>1/2$ or a unit step to the left with probability $1-q$. For large $\nu$,  ${\cal P}_\nu(x)\to\exp \left[ -(x-(2q-1)\nu)^2\big/(8q(1-q)\nu)\right]\big/\sqrt{8\pi q(1-q)\nu}$, i.e. the diffusional limit. ${\bf{\cal B}}_\mu$ and $\gamma_\mu$ are obtained by performing the Gaussian integration 
$\int_{-\infty}^{\infty}x^{\mu} {\cal P}_\nu(x)\,dx$. The return probability  for such RW is~\cite{Weiss}~\cite{ErrorWeiss}
\begin{equation}
Q_0=1-\lim_{z\to1}\frac{1}{\frac{1}{2\pi}\int_{-\pi}^{\pi}\frac{dy}{1-z[qe^{i y}+(1-q)e^{-i y}]}}=2(1-q).
\label{qobias}
\end{equation}
Eventually, from Eq.~(\ref{momentGen}), the first two moments for a biased $1$-dimensional RW are
\begin{equation}
\begin{split}
& \langle x(t) \rangle \sim \frac{1}{A\Gamma[1+\alpha]} Vt^{\alpha}
\quad,\quad
\langle x(t)^2 \rangle \sim \frac{2}{A^2\Gamma[1+2\alpha]}V^2 t^{2\alpha}
\end{split}
\label{biasmoment}
\end{equation}
where $V=2(1-q)\Big/ \left[ (2q-1)){\text {Li}}_{-\alpha}\left(2[1-q]\right) \right]$. Comparison between theoretical result and simulations of QTM is presented in Fig.~\ref{figbias}. The response to bias is nonlinear in time but also in $q$, as is seen from the form of $V$ (Fig.~\ref{figbias} {\bf{(b)}}). In the limit of $q\to 1/2$ the response in $q$ is: $V\sim (2q-1)^\alpha$, this non-linear scaling was previously predicted in~\cite{Bertin} by scaling arguments and in~\cite{MonthusSec} for very small $\alpha$. Notice also that $\langle X^2 \rangle -\langle X \rangle ^2 \sim (V /A)^2 \left[2/\Gamma[1+2\alpha]-1/\Gamma^2[1+\alpha] \right] t^{2\alpha}$ and behaves super-diffuseivily for $\alpha>1/2$. Such super-diffusive behavior has been observed in quite a few studies of disordered systems~\cite{Heuer, Franoshc, BiroliB, Binder, Lindenberg,Voituriez}.

\begin{figure}[h!]
\begin{center}

                \includegraphics[width=0.42\textwidth]{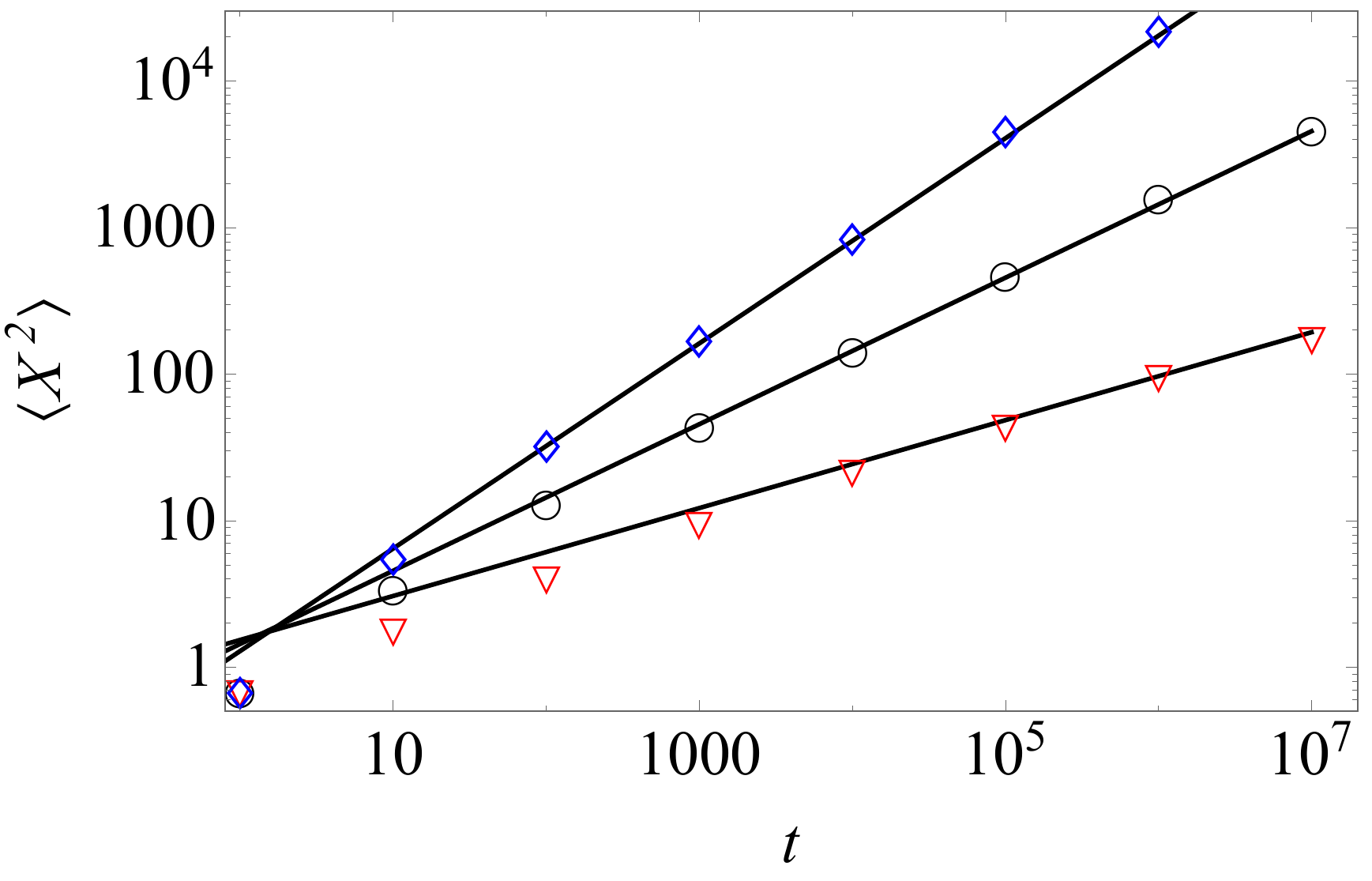}

\end{center}
\caption{ Second moment of position as function of time for unbiased QTM on a cubic lattice. Lines are analytical predictions provided by Eq.~(\ref{cubicSec}) and symbols are numerical simulations. {\color{blue}$\diamondsuit$} is $\alpha=0.7$, $\bigcirc$ is $\alpha=0.5$ and {\color{red}$\bigtriangledown$} is $\alpha=0.3$. $10^4$ realizations of disorder were used for averaging and $A=1$.
}
\label{fig3d}
\end{figure}

The second example is of a non-biased RW on a cubic lattice that can perform $6$ different unitary steps, two for every dimension. Any transition of the form ${\bf{x}}=(x,y,z)\to(x\pm1,y,z)$ has probability $1/6$ (similarly  in $y$ and $z$ directions). We again take the asymptotic limit of large number of steps and ${\cal P}_\nu({\bf{x}})\to\exp\left[-3(x^2+y^2+z^2)\big/2\nu \right]\Big/\sqrt{(2\pi\nu/3)^3}$. Due to the symmetry of the process, the first moment is strictly $0$ and the second moment is dictated by the fact that $\int_{\bf{x}} (x^2+y^2+z^2) {\cal P}_\nu({\bf{x}})\,d{\bf{x}}\sim\nu$. The return probability for a cubic lattice was already calculated in~\cite{Watson} while the analytic expression $Q_0=1-32\pi^3\Big/(\sqrt{6}\Gamma[1/24]\Gamma[5/24]\Gamma[7/24]\Gamma[11/24])\approx0.34057...$ was provided in~\cite{Glasser}. 
According to Eq.~(\ref{momentGen}) the second moment is 
\begin{equation}
\langle |{\bf{x}}|^2 \rangle \sim \frac{0.783}{A\Gamma[1+\alpha]{\text{Li}}_{-\alpha}\left(0.3405\right)}t^{\alpha},
\label{cubicSec}
\end{equation}
where we explicitly used the numerical value of $Q_0$. The comparison to simulations is presented in Fig.~\ref{fig3d}.


The presented quantitative representation of QTM in terms of CTRW (as described by Eq.~(\ref{mapPdf})) is applicable in any situation where $Q_0$ is less than $1$. Specifically this occurs for systems with dimension $>2$ or any driven system~\cite{Heuer,Lindenberg,Bechinger,Voituriez} with quenched trapping disorder.  Additionally, the mapping will be of value  for disentangling the nature of observed anomalous diffusion~\cite{BurovPnas,Meroz,IgorFelix}. While the simple temporal mapping covers a broad range of disordered systems, possible generalizations of the method are  in place. This includes  $2$-dimensional systems. Existent duality~\cite{Sollich} between trap and barrier models suggests that some variation of the mapping can be applicable to the general case of transport on random potential landscape~\cite{SokolovCamb}.

This work was partially supported by the Pazy Foundation. I thank E. Barkai for many discussions.

\bibliographystyle{apsrev4-1} 
\bibliography{./quenchedLit} 

\begin{thebibliography}{43}
\bibitem [{Kro()}]{Kronecker}%
  \BibitemOpen
  \href@noop {} {}\bibinfo {note} {$\delta_{a,b}$ is the Kronecker delta
  function}\BibitemShut {NoStop}%
\bibitem [{Err()}]{ErrorWeiss}%
  \BibitemOpen
  \href@noop {} {}\bibinfo {note} {Note that the calculation in~\cite{Weiss} is
  erroneous due to existence of additional pole}\BibitemShut {NoStop}%
\end{thebibliography}

\end{document}